\documentclass[twocolumn,prl,showpacs,amsmath,amssymb]{revtex4}

\usepackage{graphicx}
\usepackage{dcolumn}
\usepackage{bm}

\begin{document}


\title{Novel features of nuclear forces and shell evolution in exotic nuclei}
\author{Takaharu Otsuka$^{1,2}$, Toshio Suzuki$^{3}$,  
Michio Honma$^4$, Yutaka Utsuno$^5$, Naofumi Tsunoda$^1$, 
Koshiroh Tsukiyama$^1$ and Morten Hjorth-Jensen$^6$}\protect
\affiliation{$^1$Department of Physics, University of Tokyo, Hongo, Bunkyo-ku,
         Tokyo 113-0033, Japan}
\affiliation{$^2$Center for Nuclear Study, University of Tokyo, Hongo, 
         Bunkyo-ku, Tokyo 113-0033, Japan}
\affiliation{$^3$Department of Physics, Nihon University, Sakurajosui,
         Setagaya-ku, Tokyo 156-8550, Japan}
\affiliation{$^4$
Center for Mathematical Sciences, University of Aizu,
 Tsuruga, Ikki-machi, Aizu-Wakamatsu, Fukushima 965-8580, Japan}
\affiliation{$^5$Japan Atomic Energy Agency,
Tokai, Ibaraki, 319-1195 Japan}
\affiliation{$^6$Department of Physics and Center of Mathematics for 
Applications, University of Oslo, N-0316 Oslo, Norway}
\date{\today}

\begin{abstract}
\noindent
Novel simple properties of the monopole component of the effective 
nucleon-nucleon interaction are presented, leading to the so-called  
monopole-based universal interaction. 
Shell structures are shown to change as functions of $N$ and $Z$ 
consistently with experiments.  
Some key cases of this shell evolution are discussed, clarifying 
the effects of central and tensor forces.
The validity of the present tensor force is examined in terms of the 
low-momentum interaction V$_{low k}$ and the Q$_{box}$ formalism.
\end{abstract}



\maketitle

\indent
Exotic nuclei provide us with new phenomena which are not found in
stable nuclei.  One of them is the evolution of shell
structure as function of proton number ($Z$) or neutron 
number ($N$) \cite{gade,sorlin}.  
The evolution ends up, in some cases, with the appearance 
of new magic numbers and/or the disappearance of conventional ones.  
As $Z$ increases, there are more exotic isotopes between the 
$\beta$-stability line and the drip line, creating a wider frontier.
Most of such exotic nuclei are far inside the drip line, being well bound. 
The driving force behind the change in  their structure should be the
combination of the unbalanced $Z/N$ ratio and the nuclear 
force \cite{magic}.
Thus, it is crucial to see robust basic features of the
nuclear force  in exotic nuclei.  
We  present in this Letter 
novel simple properties of the monopole component of  
shell-model interactions which can reproduce experimental data.
While the shell evolution due to the tensor force has been suggested 
in \cite{tensor}, we introduce here an interaction which includes the central force 
also, moving closer to the complete picture.  

We start with selected shell-model nucleon-nucleon (NN) interactions 
which are successful in describing experimental data.
These interactions were obtained based on so-called microscopic 
interactions, derived for example with the $G$-matrix approach 
\cite{KB-G,G-mat} starting from a bare NN interaction 
and incorporating short-range repulsion and core polarization.  
In order to reproduce experimental data, however, the microscopic 
interaction has to be modified empirically, as is the case for 
the families of the USD \cite{USD}, KB3 \cite{KB} and 
GXPF1 \cite{GXPF1} interactions.  We shall take $pf$-shell 
first, and analyze the GXPF1A interaction \cite{GXPF1A} and the corresponding 
$G$-matrix interaction \cite{G-mat}.

The monopole matrix element of a given two-body interaction, $V$, is
defined as 
\begin{equation} 
v_{{\rm m;} \,j, \,j'} 
= \frac{\sum_{k,k'} \langle j k j' k' | V | j k j' k' \rangle}
{\sum_{k,k'} \,1} \, , 
\label{mono}
\end{equation}
where $j$ denotes a single-particle orbit with $k$ being its 
magnetic substate and
$\langle \, \cdot \cdot | V | \cdot \cdot \, \rangle$ is the 
antisymmetrized  two-body matrix element.  
The monopole component of $V$ is written, for $j$$\neq$$j'$, as 
${\sum_{j,j'} v_{{\rm m;} \,j, \,j'} n_{j} n_{j'}}$, 
where $n_{j}$ is the occupation number (operator) of orbit $j$ 
\cite{mononn}. 
The monopole component is nothing but the average over all orientations.  
It was introduced by Bansal and French \cite{bansal}, while its 
relevance to the effective shell model interaction was discussed by 
Poves and Zuker \cite{KB}.
Recently, the monopole component of the spin-isospin interaction has 
been shown to modify even the magic structure in exotic nuclei 
\cite{magic}, and the specific and substantial role of the tensor 
force was shown in Ref.~\cite{tensor}.  
Note that $v_{{\rm m;} \,j, \,j'}$ is defined either with   
isospin, $T$=0 or 1, or in the proton-neutron scheme, while    
``$j,j'$'' may be omitted for brevity.

\begin{figure*}[tb]
 \begin{center}
  \includegraphics[width=13.8cm,clip]{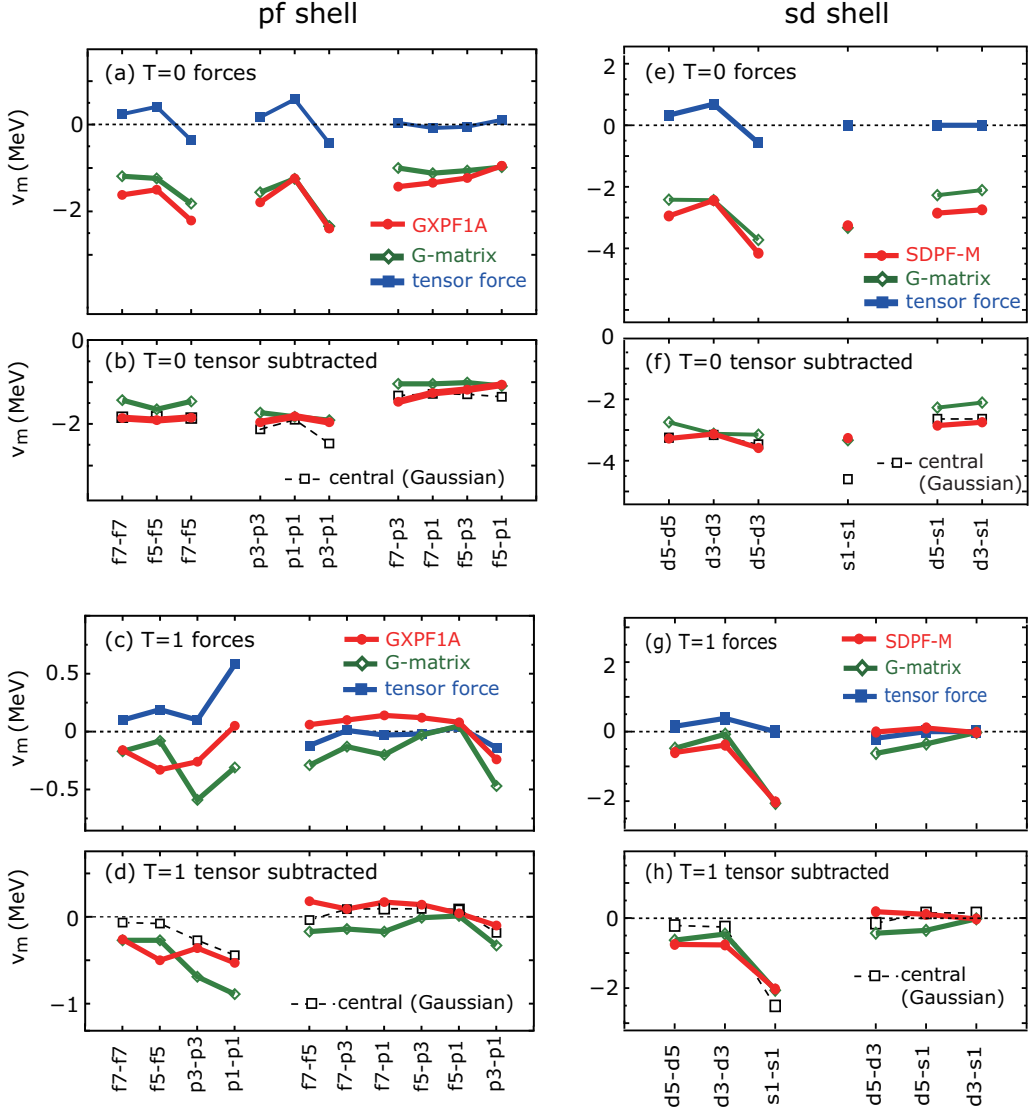}
 \caption{(color online) Monopole matrix elements of various forces for 
  (a-d) $pf$ and (e-h) $sd$ shells.  In (b,d,f,h), the tensor-force 
  effect is subtracted from the others, and results from a Gaussian 
  central force are shown.
}
  \label{pfsd_mono}
 \end{center}
\end{figure*}

The importance of the monopole interaction for exotic nuclei originates
in its linearity.  As the orbit $j'$ is occupied, the single-particle 
energy (SPE) of an orbit $j$, $\epsilon_j$, is changed by
\cite{mononn},    
\begin{equation} 
\Delta \epsilon_{j} \, 
= \, v_{{\rm m;} \,j, \,j'}  n_{j'}.  
\label{linear}
\end{equation}
For $j'$=$g_{9/2}$ as an example, $n_{j'}$ takes values up to 10.  
Thus, the effect of the monopole component can be magnified considerably.  
By moving along the nuclear chart, one can indeed change a particular  
$n_{j'}$ substantially.
This highlights the physics of exotic nuclei compared to
that of stable nuclei, and it is of keen and urgent interest to 
clarify general and robust features of the monopole interaction. 
At shell closures, the monopole component produces effects according to
Eq.~(\ref{linear}), whereas effects of other multipole components vanish.  
The monopole component governs (spherical) 
SPEs on top of closed (sub-)shells.  In open shell systems, its effects can 
be viewed through Eq.~(\ref{linear}) as effective SPEs.
As the surface deformation with low excitation energies is 
a Jahn-Teller effect, the SPEs are crucial for collectivity too.  

Figure \ref{pfsd_mono} (a) shows $v_{{\rm m;} \,j, \,j'}$
for isospin $T$=0 
from the GXPF1A interaction, the $G$-matrix interaction \cite{G-mat} and the tensor force in the $pf$-shell.  
The tensor force refers hereafter to the $\pi$+$\rho$ meson exchange
force used in \cite{tensor}.  
The orbits ($j, j'$) are grouped as ($f$, $f$), ($p$, $p$) and ($f$, $p$).
In Fig. \ref{pfsd_mono} (a), we find two distinct kinks in the 
tensor-force values for the ($f$, $f$) and the ($p$, $p$) groups, and the
same kinks appear also in the GXPF1A and the $G$-matrix results. 
Note that each kink is a consequence of the general rule suggested in 
\cite{tensor}.
The similarities are remarkable. To shed more light on this, in 
Fig. \ref{pfsd_mono} (b)
we subtract the tensor-force contribution from the GXPF1A and the $G$-matrix
values.   This results in almost flat curves.
The ($f$, $f$) and ($p$, $p$) cases show almost the same values, 
while the ($f$, $p$) shows higher but still nearly flat values.
This can be understood in terms of radial integral of the central 
force: in the former case the radial wave functions are the same 
between $j$ and $j'$, 
while they are different in the latter.  
The flatness suggests a longer-range central force.
In order to incorporate these features, we introduce a central 
Gaussian interaction as
\begin{equation} 
V_c \, 
= \, {\sum_{S,T} \, f_{S,T} P_{S,T} \exp {(-(r/\mu)^2)}},
\label{gaussian}
\end{equation}
where $S (T)$ means spin (isospin), $P$ denotes the projection 
operator onto the channels ($S,T$) with strength $f$, and $r$ and 
$\mu$ are the internucleon distance and Gaussian parameter,
respectively.  Figure \ref{pfsd_mono} (b) shows 
results obtained by $f_{0,0}$=$f_{1,0}$=166 MeV and $\mu$=1.0 fm.  
The agreement with GXPF1A is remarkable, considering the simplicity of 
the model.
Thus, we can describe the monopole component by two simple terms: 
the tensor force generates ``local'' variations, while the Gaussian 
central force produces a flat ``global'' contribution.  
It is worth mentioning that $\mu$=1.0 fm is reasonable from 
the viewpoint of NN interaction, and deviations
from it, including the zero-range limit, worsen the agreement.  

Figure~\ref{pfsd_mono} (c) shows $v_{\rm m}$'s for $T$=1.
They are grouped for pairs of $j$=$j'$ and the rest.
The former corresponds to the standard BCS-type pairing cases.   
We first stress that the basic scale is quite different
between $T$=0 and 1: $v_{\rm m}$'s of GXPF1A are in the range  
-2.5 $\sim$ -1 MeV 
for $T$=0, whereas for $T=1$ they are in the range -0.3 $\sim$ 0.2 MeV.  
The sharp rise for $j$=$j'$=$p_{1/2}$ occurs in all three 
interactions as a characteristic fingerprint of the tensor force. 
Note that $v_{\rm m}$'s for the GXPF1A interaction ($G$-matrix) are 
mostly repulsive (attractive) for $j$$\neq$$j'$.  
A similar repulsive correction to $G$-matrix values occurs in the $sd$-shell 
as well.
We subtract the tensor contribution 
as was done in Fig.~\ref{pfsd_mono} (b), 
and show the result in Fig.~\ref{pfsd_mono} (d) as well as those 
of the Gaussian central force with 
$f_{0,1}$=0.6$f_{0,0}$ and $f_{1,1}$=-0.8$f_{0,0}$.  
The basic feature can be reproduced, apart from some deviations in 
the ($f$,$f$) cases, which may indicate stronger pairing correlations.

\begin{figure}[tb]
 \begin{center}
  \includegraphics[width=5.8cm,clip]{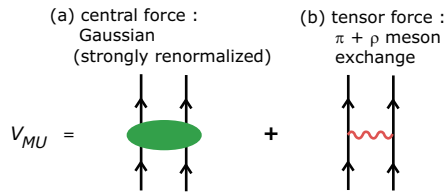}
\caption{(color online) Diagrams for the $V_{MU}$ interaction.
}
  \label{diagram}
 \end{center}
\end{figure}

Figure~\ref{pfsd_mono} (e-h) exhibits $v_{\rm m}$'s 
in the $sd$-shell,  similar to what is shown in Figs.~\ref{pfsd_mono} (a-d).
The SDPF-M interaction \cite{SDPF-M} is taken as the realistic 
interaction.
All features discussed for the $pf$-shell are seen, and the 
tensor-subtracted values are reproduced by the same Gaussian
central force even better for $T$=1.  One sees repulsive 
corrections to $v_{\rm m}$'s from the  $G$-matrix for $T$=1 and $j$$\neq$$j'$, 
similar to our findings in the $pf$ shell.   This correction is linked with the oxygen 
drip line, its origin has been a puzzle, 
but has recently been resolved \cite{3NF}. 

Based on the above results, 
we introduce the {\it monopole-based universal interaction}, 
$V_{MU}$.  As shown in Fig.~\ref{diagram}, $V_{MU}$ consists of
two terms.  The first term is the Gaussian central force discussed so far, 
and should contain many complicated processes including multiple 
meson exchanges.  The second one is the tensor 
force comprised of $\pi$ and $\rho$ meson exchanges \cite{tensor}.
The $V_{MU}$ interaction resembles Weinberg's original model for
Chiral Perturbation theory\cite{weinberg}, if one replaces 
Fig.~\ref{diagram}
(a) by contact terms and (b) by the one-$\pi$ exchange potential.

\begin{figure*}[tb]
 \begin{center}
  \includegraphics[width=0.82\textwidth,clip]{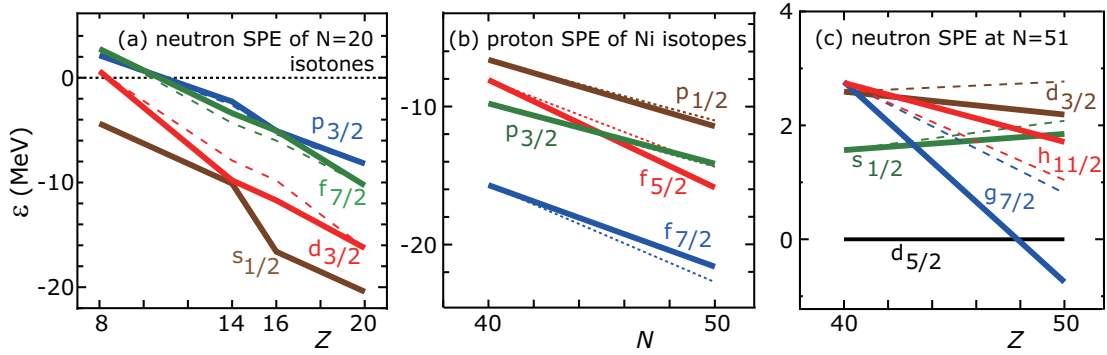}
 \caption{(color online) Single-particle energies (SPEs) calculated by $V_{MU}$ 
  interaction.  The dashed lines are obtained by the central force 
  only, while the solid lines include both the central force and the tensor force.
}
  \label{spe}
 \end{center}
\end{figure*}

Figure~\ref{spe} shows applications of $V_{MU}$ 
to the shell evolution assuming a filling configuration.
Figure~\ref{spe} (a) depicts neutron SPEs
around $N$=20 for $Z$=8$\sim$20.  Starting 
from SDPF-M SPEs at $Z$=8, one sees the evolution of the $N$=20 gap, 
in a basically consistent manner with other shell-model
studies \cite{SDPF-M,SDPF-U}. 
While the change is monotonic without the tensor force, 
the tensor force produces a sharp widening from $Z$=8 to 14, 
and then stabilizes the gap 
towards $Z$=20.  
It is worth mentioning that the normal SPEs arise at $Z$=20, 
whereas at $Z$=8 the inversion between $f_{7/2}$ and $p_{3/2}$ occurs
and $d_{3/2}$ is rather close to $p_{3/2}$, leaving the major
gap at $N$=16.
The central force lowers the neutron $d_{3/2}$ SPE more than the $f_{7/2}$ SPE
as protons occupy the  $sd$-shell due to larger overlaps, 
yielding a wide $N$=20 gap at $^{40}$Ca.
The $N$=20 gap at $Z$$\sim$14 is, however, largely due to the tensor force, 
and becomes smaller if protons are excited to $d_{3/2}$.

Figure~\ref{spe} (b) shows proton SPEs for $^{68-78}$Ni starting 
from empirical values \cite{Grawe} at $N$=40.  The SPE of $p_{1/2}$
is not known empirically, and is placed above $p_{3/2}$ by the
energy difference predicted by the GXPF1A interaction.
The orbit $f_{5/2}$ crosses $p_{3/2}$ at $N$=45 consistently with 
a recent experiment \cite{Cu}, and the $f_{7/2}$-$f_{5/2}$ 
splitting is reduced by 2 MeV from $N$=40 to 50.
For both, the tensor force plays crucial roles.
Apart from certain differences, the trend is seen in other 
shell model results, {\it e.g.}, \cite{Heyde}.

Figure~\ref{spe} (c) shows neutron SPEs relative to 
$d_{5/2}$ on top of  
$^{90}$Zr-$^{100}$Sn, starting from empirical values at $Z$=40
obtained by averaging with spectroscopic factors \cite{ENSDF}.
The lowering of $g_{7/2}$ is remarkable \cite{FP}.  If there were no 
tensor-force effects, $g_{7/2}$ and $h_{11/2}$ do not repel, ending  
up with quite a different shell structure for $^{100}$Sn, making 
this nucleus much softer.  The closer spacing of $g_{7/2}$ and $d_{5/2}$
in $^{101}$Sn seems to be seen experimentally \cite{101Sn}.  

\begin{figure}[tb]
 \begin{center}
  \includegraphics[width=6.5cm,clip]{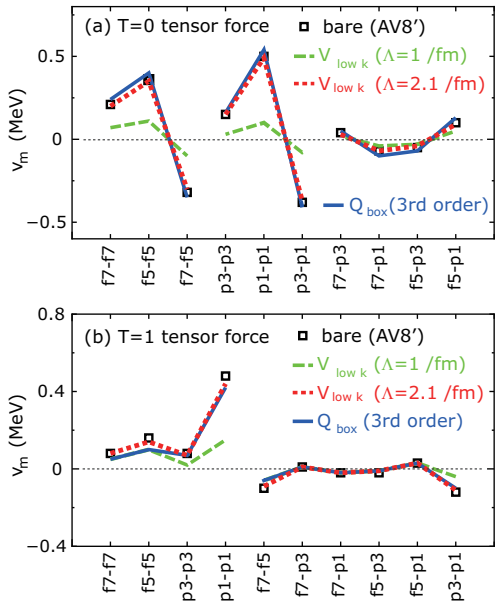}
 \caption{(color online) Tensor forces in AV8' interacton, in low-momentum interactions
obtained from AV8', and in the 3rd-order Q$_{box}$ interaction for (a) T=0 and
(b) T=1.
}
  \label{VlQb}
 \end{center}
\end{figure}

We now discuss whether the simple tensor force in $V_{MU}$ can be 
explained microscopically or not.  We take the AV8' interaction \cite{AV8'}
and examine how the tensor force obtained by the spin-tensor 
decomposition changes in the following processes.
We derive a  low-momentum interaction V$_{low k}$ \cite{Vlowk} and  
calculate $v_{\rm m}$'s,  
as shown in Fig.~\ref{VlQb} (a,b) for $T$=0, 1, varying the 
cut-off parameter, $\Lambda$.  For the usual value $\Lambda$=2.1 (fm$^{-1}$), 
the result is very close to the bare AV8' tensor force contribution.  
We then perform the 3rd-order Q-box calculation with folded 
diagram corrections \cite{G-mat}, 
in order to include medium effects like core polarization.
The result still resembles $v_{\rm m}$'s of the bare tensor part. 
Thus we can confirm that the treatments of the short-range correlation 
and the medium effects do not change much $v_{\rm m}$'s 
of the tensor force.  This near-independence may be interpreted 
in terms of specific and complicated angular momentum coupling in
the tensor force.  For instance, the 2nd-order perturbation by two tensor forces 
yield mainly a central force. 
For unusual values like $\Lambda$=1 (fm$^{-1}$),
deviations arise, as expected.  

The central force depends strongly on $\Lambda$.   
For $\Lambda$=2.1 (fm$^{-1}$), $v_{\rm m}$'s of the central part of 
$V_{low k}$ are scattered around the values of 
Fig.~\ref{pfsd_mono} (b).  
This result is promising, but more studies are needed.

In summary, we have presented novel general properties of the monopole
interactions, and  introduced the $V_{MU}$ interaction consisting of
simple central and tensor forces.
The persistency of the bare tensor force is examined by the V$_{low k}$
and Q$_{box}$ formalisms.
The $V_{MU}$ produces a variety of the shell evolution, connecting
stable and exotic nuclei, for instance, exotic Ne-Mg with $^{40}$Ca,
$^{68}$Ni with exotic $^{78}$Ni, and $^{90}$Zr with exotic $^{100}$Sn.
The shell structure appears to vary considerably in exotic
nuclei.  
As $V_{MU}$ has been introduced based on monopole properties, 
tests of its validity by full shell-model calculations and 
possible refinements including multipole components are of great
interest, as well as more predictions by $V_{MU}$.

We thank Dr. H.~Grawe for valuable discussions.
This work was supported in part by a Grant-in-Aid for Specially Promoted
Research (13002001).  
This work has been supported by the JSPS Core-to-Core Prgram EFES, 
and is a part of the RIKEN-CNS joint research project on
large-scale nuclear-structure calculations.


\begin{thebibliography}{}
\bibitem{mononn} For $j$=$j'$ for identical nucleons, slightly different 
                 forms are used.
\end{thebibliography}
\end{document}